\documentclass[runningheads]{llncs}
\usepackage[T1]{fontenc}
\usepackage{graphicx}
\usepackage{xcolor}     
\usepackage{colortbl}
\usepackage{threeparttable}
\usepackage{makecell}
\usepackage{booktabs, tabularx, array, enumitem}
\usepackage{caption}
\usepackage{arydshln}
\usepackage{multirow}

\begin{document}
\title{Three Years with Classroom AI in Introductory Programming: Shifts in Student Awareness, Interaction, and Performance}
\titlerunning{Three Years with Classroom AI in Introductory Programming}

\author{
Boxuan Ma\inst{1}
\and
Huiyong Li\inst{1}
\and
Gen Li\inst{1}
\and
Li Chen\inst{2}
\and
Cheng Tang\inst{1}
\and \\
Atsushi Shimada\inst{1}
\and
Shin'ichi Konomi\inst{1}}
\authorrunning{Ma et al.}

\institute{Kyushu University, Fukuoka, Japan  \\
\and
Osaka Kyoiku University, Osaka, Japan
}


%
\maketitle              % typeset the header of the contribution
\begin{abstract}
Generative AI (GenAI) tools such as ChatGPT now provide novice programmers with instant, personalized support and are reshaping computing education. While a growing body of work examines AI’s immediate impacts, longitudinal evidence remains limited on how students’ awareness, student–AI interaction patterns, and course outcomes evolve as AI becomes routine in classrooms. To address this gap, we investigate an introductory Python course across three successive AI-supported cohorts (2023–2025). Using questionnaires, coded student–AI dialogue logs, and course assessment records, we examine cohort-to-cohort shifts in students’ AI awareness, interaction practices, and learning outcomes. We find that students’ relationships with GenAI change systematically over time: familiarity and uptake become increasingly normative, and help-seeking practices evolve alongside growing AI literacy and shifting expectations of what the assistant should provide. These changes suggest that, in the AI era, the central instructional challenge is less about whether students use AI and more about how courses redefine productive learning practices while maintaining student agency. Our study offers longitudinal evidence and practical implications for designing and integrating AI programming support in course settings.

\end{abstract}
\section{Introduction}

Generative AI (GenAI) tools such as ChatGPT and AI-powered coding assistants are rapidly reshaping introductory programming education by making explanations, examples, and debugging guidance available on demand \cite{yilmaz2023augmented,kazemitabaar2024codeaid}. This immediacy can narrow a long-standing instructional gap between when novices encounter difficulties and when they receive help, especially outside lecture time or office hours. At the same time, this capability raises pedagogical concerns: because GenAI can often produce working solutions with minimal prompting \cite{denny2023conversing}, learners may shift from deliberate problem solving toward solution seeking, potentially reducing opportunities for critical thinking and reasoning through the underlying concepts \cite{ma2025scaffolding,darvishi2024impact,fan2025beware}.

While a growing body of work examines AI’s immediate impacts, longitudinal evidence remains limited on how students’ awareness of AI, student--AI interaction patterns, and course outcomes evolve as AI becomes routine in classrooms. Most existing studies rely on short-term deployments, making it difficult to separate novelty effects from stable shifts in students’ student--AI interaction practices. As a result, we still lack longitudinal classroom evidence that jointly tracks (1) how students enter courses with different prior familiarity and expectations, (2) how their student--AI interaction patterns shift in purpose and workflow, and (3) whether these shifts coincide with changes in course outcomes under same instructional settings.

To address this gap, we conduct a three-year classroom study in an introductory Python course spanning three AI-enabled cohorts (2023--2025). We triangulate questionnaires on students' familiarity with AI, usage, and perceived benefit for learning programming; coded student--AI dialogue logs that capture the distribution and sequencing of prompt types; and course records, including weekly assignments and final grades.

We investigate three research questions:
\begin{itemize}[label=\textbullet]
\item \textbf{RQ1 (Awareness):} How did students’ awareness and familiarity with AI change across cohorts?
\item \textbf{RQ2 (Interaction):} How did student--AI interaction shift across cohorts in terms of prompt types and their sequential workflows?
\item \textbf{RQ3 (Performance):} Relative to the pre-AI baseline, how did weekly assignment performance and final course outcomes change over the three years?
\end{itemize}

\section{Related Work}

GenAI is reshaping programming education by lowering barriers to help-seeking and automating a broad range of programming tasks, including code generation, debugging, code explanation, and the delivery of personalized feedback \cite{denny2023conversing}. Early research in this area has primarily focused on evaluating the capabilities of GenAIs on programming tasks \cite{phung2023generative,sarsa2022automatic,savelka2023thrilled}, showing that they are able to address common programming challenges with impressive results. A growing body of work also focuses on the human side. Studies have systematically examine how students perceive and experience the use of GenAI in programming education \cite{yilmaz2023augmented,shoufan2023exploring,ma2024enhancing,ma2024exploring}. This is critical because the effectiveness of GenAI as educational tools depends not only on what they can do, but also on how learners and teachers engage with them and adapt their practices accordingly. Studies consistently show that students view GenAI as valuable supports in programming education \cite{yilmaz2023augmented,shoufan2023exploring,ma2024enhancing}. They appreciate the immediacy and accessibility of GenAI, emphasizing how quickly it can provide solutions or guidance compared to searching online resources or waiting for human assistance \cite{yilmaz2023augmented}. Beyond efficiency, students also highlight the clarity of AI-generated feedback, which helps them understand concepts and debug code more effectively \cite{ma2024enhancing}. At the same time, students remain cautious: many acknowledge that answers can be incorrect or misleading \cite{shoufan2023exploring,ma2024enhancing}. They also worried about the unintended consequences of unmoderated GenAI use lead to superficial understanding and lower knowledge retention \cite{skjuve2023user,tlili2023if,ma2025generative}. In the absence of deliberate practice, these behaviors risk undermining the development of problem-solving persistence, debugging proficiency, and critical thinking \cite{kasneci2023chatgpt,pankiewicz2023large,shoufan2023exploring}. 

Although GenAI are seen as offering unprecedented scalability and personalization, empirical studies highlight the need for educators to remain abreast of technological developments while guiding students on ethical use \cite{prather2023robots}. Some advocate restricting or banning GenAI use to preserve academic integrity, whereas others experiment with integration strategies that leverage AI for formative support while safeguarding assessment integrity \cite{becker2023programming}.

Overall, prior work establishes both the promise of GenAI for programming support and practical concerns around reliability, overreliance, and assessment integrity. However, much of the evidence to date is either capability-focused—emphasizing what AI can do—or based on short instructional windows (e.g., a week or a single term). As GenAI rapidly diffuses into students’ everyday practices, an open question is how classroom use evolves once GenAI becomes “normal” rather than novel. To address this gap, we present a three-year classroom study that triangulates questionnaires, coded dialogue logs, and course records to characterize longitudinal changes in student awareness, interaction patterns, and course outcomes across a pre-GenAI baseline and three AI-supported cohorts.

\section{Method}

We conducted a three-year study in an introductory undergraduate Python programming course at a national university in Japan. The study spanned from 2023 to 2025 and consisted of three offerings of the same course, with a total of 248 students. Throughout each course, all students were given access to an AI tool as a supplementary learning resource. Access was provided through a custom, web-based interface that enabled real-time interaction with GPT-based AI. All conversations between students and AI were systematically logged for subsequent analysis. The study adopted a mixed-methods approach, combining quantitative and qualitative data collection. The study received approval from the university’s ethics review board prior to the start of the study.

\subsection{Course Structure}

This first-year undergraduate course, designed for beginners, covered the foundations of the Python programming language in 14 lessons delivered over one semester. Each 90-minute lesson was divided into two segments: the first 45 minutes were dedicated to direct instruction, supported by lecture presentations and slides, followed by 45 minutes of hands-on programming tasks related to the lesson’s topic (e.g., write a specific function based on the requirements). This structure was intended to enable students to apply the theoretical concepts introduced in class. 

Two instructors delivered the three course offerings, and the curriculum was largely unchanged between 2023 and 2025, which allowed us to maintain consistency across offerings. In each lesson, students completed programming exercises. Each course included a total of 57 exercises (averaging 4.75 per lesson). All exercises were submitted through the university’s Learning Management System (LMS), which automatically logged and evaluated each exercise submission. Students had access to a variety of learning resources beyond AI, including lecture slides and support from the course instructors, and were instructed to cite any external sources used in their work.

\subsection{Participants}

The study involved a total of 248 undergraduate students over a three-year period (62 participants in 2023, 126 in 2024, and 60 in 2025), across three offerings of the same course in the university. Participation was voluntary, informed consent was obtained from all participants, and no aspect of participation affected students’ course grades. The distribution of demographic characteristics was generally consistent across years. Overall, the majority of participants (96.8\%) were first-year students at the time of participation, with a near-equal gender distribution (53.2\% male, 45.8\% female, and 1\% identified as non-binary). While most participants were domestic students from Japan, approximately 14.5\% were international students from diverse countries. Most participants (90\%) self-identified as beginners in Python programming, and this proportion remained relatively consistent across all three years of the study. 

\subsection{Data Sources}
To gather a comprehensive understanding of student interactions with AI and student perceptions of AI, we employed a multifaceted data collection approach by collecting questionnaire from students and student-AI interaction logs.

\textbf{Questionnaire.} At the beginning of the course, all participants completed a questionnaire capturing demographics (e.g., major and gender), self-reported Python proficiency, and prior GenAI experience. Specifically, the questionnaire asked about students’ familiarity with GenAI such as ChatGPT, whether and how they currently use AI tools, their perceived benefit of using AI for learning Python, and the ways they expected AI could support programming learning. There is also an open-ended question inviting students to describe GenAI in their own words. Detailed questionnaire items are provided in the supplementary material\footnote{https://bit.ly/3Mi25sg \label{appendix}}.

\textbf{Dialogue Logs} A primary data source for understanding students' usage patterns and AI's response was student-AI dialogue logs. Dialogue data has become an important data source to understand programming experiences and coding approaches, particularly when interacting with AI \cite{kazemitabaar2023studying}. For each question asked by students, we closely examined its content and AI generated responses through a thematic analysis. 

%\textbf{Automated Assessment Submissions} The exercises were delivered and assessed using a Moodle plugin for automated programming assessment. Students could read the problem statement, input their code, and submit their solutions to be automatically checked against pre-defined test cases. Each submission were logged and the scores are recorded automaticly.

\subsection{Data Analysis}

We used a mixed-methods design aligned with our three research questions. RQ1 (Awareness) was examined through longitudinal comparisons of survey measures across cohorts. RQ2 (Interaction) was examined through thematic coding of student--AI dialogue logs, followed by quantifying code distributions and transitions across years. RQ3 (Performance) compared cohort-level outcomes on weekly assignments and final grades. To provide a pre-GenAI point of reference for RQ3, we also include a separate baseline offering taught by the same instructor in early 2023 (n=39), when students in our context had not yet widely adopted GenAI. We compare this baseline against three subsequent GenAI-enabled offerings (2023--2025).

\paragraph{Thematic Coding.}
We conducted a thematic analysis of the dialogue logs. From 10,632 recorded interactions, we performed stratified random sampling across cohorts and selected 2,782 instances (26\%) for manual coding. Each cohort contributed a proportionate share of the sample to ensure balanced representation.

Drawing on coding schemes from prior studies \cite{kazemitabaar2024codeaid,ma2024exploring}, we organized the analysis around two high-level dimensions: student prompts and AI-generated responses. For student prompts, we characterized the types of questions and requests students made, following established codebooks for programming-related student--AI interactions \cite{silva2024learning}. For AI-generated responses, we assessed pedagogical quality and course relevance. Building on categories adapted from prior work \cite{kazemitabaar2024codeaid,ma2025scaffolding}, we coded responses along two independent axes: \emph{Correctness} and \emph{Helpfulness}. Correctness captures whether the response is technically accurate or contains any errors, while Helpfulness captures whether it enables the student to make progress. A response can be correct but not helpful (e.g., if it is irrelevant or outside the course scope). Detailed code definitions and descriptions are provided in the supplementary material\footref{appendix}.

The coding procedure was identical across both dimensions. Two researchers jointly reviewed an initial set of 800 randomly sampled interactions, generating preliminary sub-dimensions and codes. They then independently coded an additional 150 samples using this preliminary codebook. Then, the results were discussed with the course instructors, and disagreements were resolved through consensus, leading to a refined version of the codebook. To establish reliability, the two researchers independently coded 334 further samples, and inter-rater agreement was calculated using Cohen’s Kappa and percentage agreement. Once reliability thresholds were satisfied, the researchers proceeded to independently code an additional 2,448 interactions randomly drawn from the remaining dataset (averaging across sub-dimensions, Cohen's $\kappa$ was .88 for student-prompt coding and .91 for AI-response coding).

% \cite{miles1994qualitative,neuendorf2017content}

\begin{figure}[tb]
\centering
\includegraphics[scale=0.275]{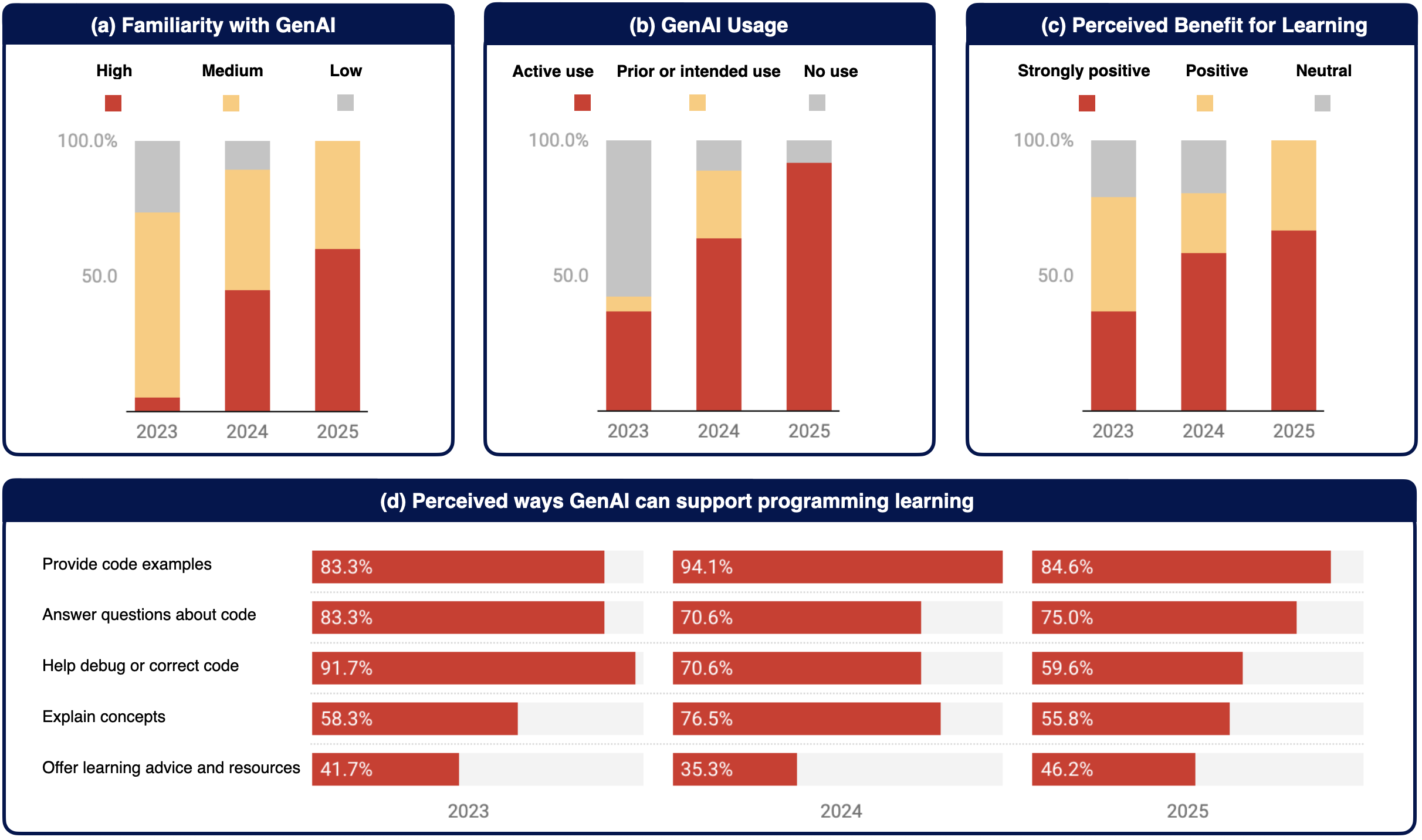}
\caption{Shifts in students’ GenAI Perspectives and Awareness. (a) Familiarity with GenAI, (b) GenAI usage, and (c) perceived benefit for learning. (d) Perceived ways GenAI can support programming learning.}
\label{survey}
\end{figure}

\section{Results}

\subsection{Shifts in Student Perspectives and Awareness (RQ1)}

%
%Response options were collapsed into three levels for readability: for Q1, (High = ‘can explain everything/some extent’, Medium = ‘somewhat understand’, Low = ‘don’t understand well/don’t know at all’); for Q2, (Active use, Former-or-Intended use, No use); for Q3, (Positive, Neutral, Negative). Bars show within-cohort percentages.

\subsubsection{Familiarity with GenAI}
As shown in Fig.~\ref{survey}(a), students’ self-reported familiarity with GenAI increased substantially across cohorts. In 2023, only 5.2\% of respondents reported high familiarity, while the majority fell into the medium category (68.4\%) and over a quarter reported low familiarity (26.3\%). By 2024, the distribution shifted upward, with nearly half reporting high familiarity (47.2\%), a comparable share reporting medium familiarity (47.2\%), and only 5.6\% remaining in the low category. By 2025, all respondents reported at least medium familiarity, and 60\% reported high familiarity, suggesting that GenAI had become a more common and better understood tool before entering the course.

\subsubsection{GenAI Usage}
Fig.~\ref{survey}(b) shows a parallel rise in GenAI usage. In 2023, only 36.8\% reported active use, and a substantial portion reported no use. By 2024, active use increased markedly (63.9\%), and the share of students reporting no use dropped to 11.1\%. By 2025, GenAI use became nearly universal: 91.7\% reported active use and only 8.3\% reported no use. Overall, the cohort-to-cohort pattern suggests that GenAI shifted from an optional tool used by a minority to a routine resource for most students.

\subsubsection{Perceived benefit for learning}
Students’ perceived benefits of GenAI were positive across all three years and increased over time (Fig.~\ref{survey}(c)). In 2023, responses were split between positive, strongly positive and neutral, with relatively fewer students expressing strongly positive views. By 2024, positive perceptions strengthened, with neutral responses decreasing. By 2025, perceptions were uniformly favorable: 100\% of respondents rated GenAI as positive or strongly positive for learning, and negative responses were absent across cohorts. This suggests growing confidence in GenAI as a learning aid as students’ prior exposure increased. The upward trend may also reflect concurrent improvements in GenAI models and tools, alongside better integration into students’ learning practices.

\subsubsection{Perceived ways GenAI can support programming learning}
Fig.~\ref{survey}(d) shows that students primarily viewed GenAI as a code-oriented support tool across cohorts, most frequently selecting code examples and code-related Q\&A. The only notable cross-year shift was a decline in debugging/correction expectations, dropping from 91.7\% (2023) to 59.6\% (2025). Overall, these responses suggest that students increasingly conceptualized GenAI as a general programming helper rather than a single-purpose debugging tool.

\begin{figure}[t]
\centering
\includegraphics[width=\textwidth]{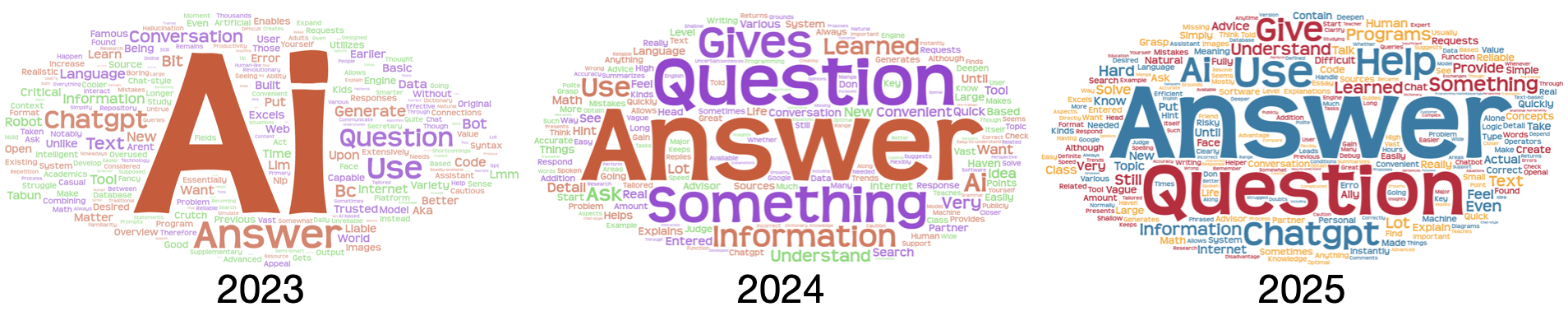}
\caption{Word clouds of students’ GenAI awareness (larger words indicate higher frequency in students’ responses).}
\label{fig:awareness}
\end{figure}

 \subsubsection{Conceptual Shift in How Students Frame GenAI}

In our questionnaire, students were invited to describe AI in their own words. The resulting descriptions were compiled to generate the word clouds shown in Fig. 2, which visualize changes in the most frequently used terms. 

In 2023, students’ awareness was dominated by broad, high-level terms such as “AI,” “ChatGPT,” and “Conversation”. These responses reflected an early-stage understanding that focused on novelty and general potential rather than concrete functions. By 2024, students’ pre-class awareness had become more comprehensive and detailed, with frequent references to “Answer,” “Question,” and “Information.” 2025, however, their vocabulary became more educationally oriented, including words such as “Help,” “Understand,” and “Learned.” This reflects a growing tendency to view AI not only as answer providers but as learning partners that can support understanding and reflective engagement. This shift suggests that students entered the course with a more grounded and multifaceted awareness of AI’ functions, shaped by their wider exposure to AI tools in everyday contexts.

\begin{figure}[t]
\centering
\includegraphics[width=\textwidth]{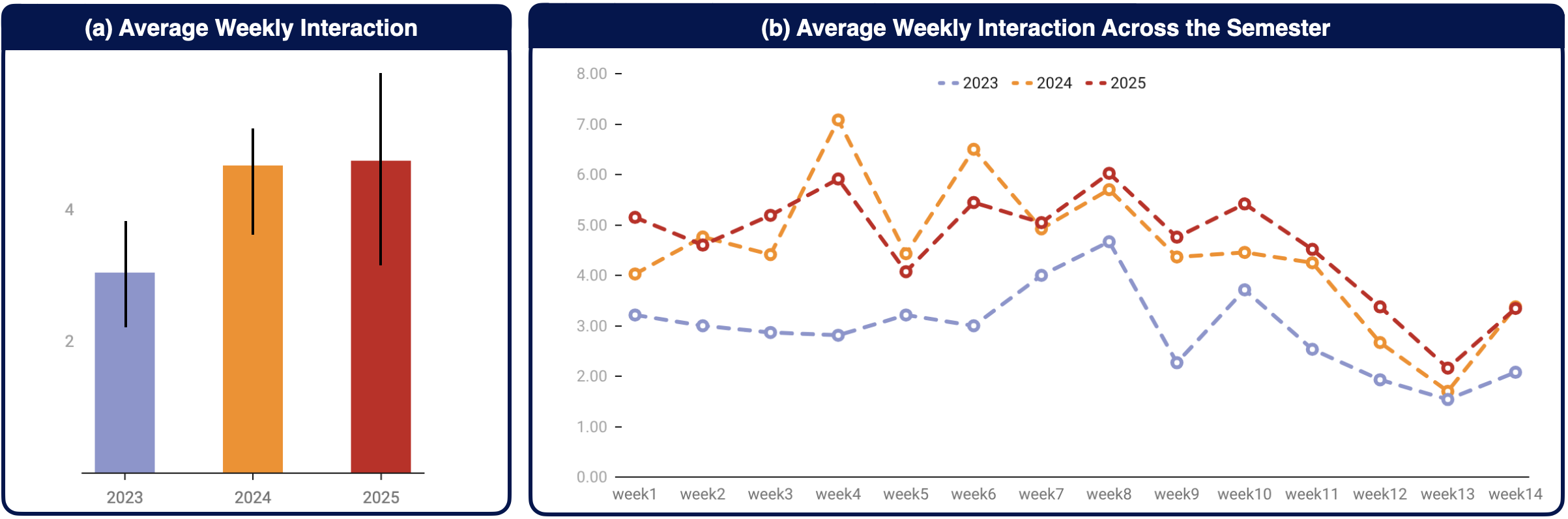}
\caption{Weekly trends in average conversation depth.}
\label{weekly}
\end{figure}

\subsection{Shift of Interactions (RQ2)}

\subsubsection{Interaction Frequency}

We first examined how frequently students interacted with GenAI during weekly coursework. We operationalized interaction frequency as the number of student requests per student-week, where a request is a student message directed to the GenAI assistant (Fig.~\ref{weekly}(a)). Overall, requests increased substantially after 2023: the mean requests per student--week rose from 3.05 in 2023 to 4.68 in 2024 and remained comparable in 2025 (4.75). This cohort difference was statistically significant (Kruskal--Wallis $H=31.91$, $p<0.0001$), driven by the gap between 2023 and both later cohorts; the 2024 and 2025 cohorts did not differ significantly. 

Fig.~\ref{weekly}(b) further highlights clear cohort differences in week-by-week interaction: throughout the semester, students in 2023 consistently made fewer requests per week than those in 2024 and 2025. The trajectories also suggest strong coupling with the curriculum, with peaks occurring in weeks that introduced more conceptually demanding topics and multi-step programming tasks, and lower activity in the first and last weeks when exercises were more foundational or focused on wrap-up.

The lower interaction levels in 2023 may plausibly reflect early-stage GenAI literacy. Many students were still learning how to formulate effective prompts and how to refine them when initial responses were incomplete or unhelpful, which may have increased friction and frustration, leading some students to disengage from iterative help seeking. By 2024 and 2025, GenAI had become a more routine learning tool (also reflected in the survey), and students more often continued seeking help via follow-up requests and iterative refinements. Taken together, the results suggest that cohort-level growth in GenAI familiarity coincided with more frequent help-seeking, while week-to-week variation continued to track task difficulty and curricular content.

\begin{figure}[t]
\centering
\includegraphics[width=\textwidth]{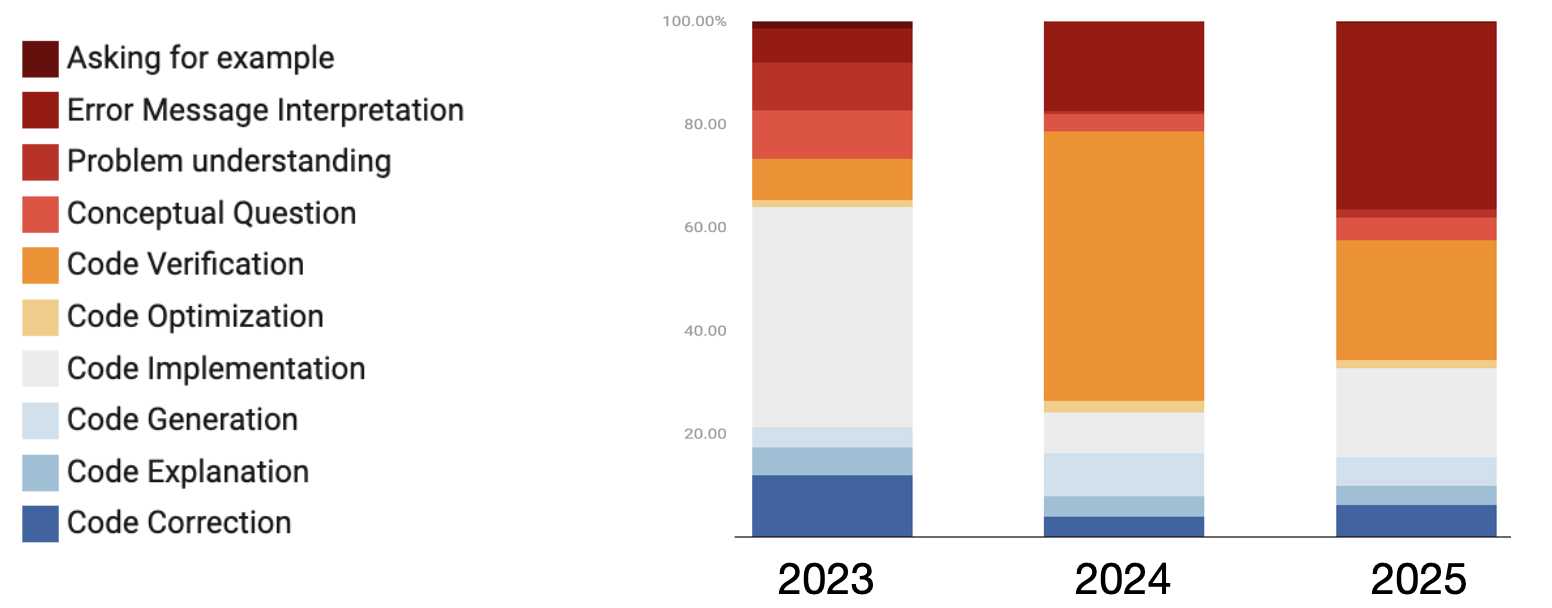}
\caption{Distribution of student prompt types across years (100\% stacked).}
\label{prompt_type_distribution}
\end{figure}

\subsubsection{Student Prompt Types}

To examine how students’ help-seeking behaviors evolved across the three years of AI-supported instruction, we compared the distribution of coded student prompt types in 2023, 2024, and 2025. The results reveal a clear shift in students’ interaction over time.

In 2023, students’ prompts were dominated by Code Implementation prompts (42.67\%), indicating an implementation-driven pattern in which learners primarily asked the AI to help write or complete code. In contrast, prompts related to Code Verification (8.00\%) and Error Message Interpretation (6.67\%) were relatively infrequent. In 2024, this pattern changed markedly: Code Verification became the most prevalent prompt type (52.25\%), while Code Implementation dropped to 7.87\%. This suggests that students increasingly used the AI to check, validate, and confirm their own solutions rather than to generate implementations directly. By 2025, students’ help-seeking behaviors shifted again toward debugging-oriented use. Error Message Interpretation rose to 36.18\%, while Code Verification (23.27\%) and Code Implementation (17.28\%) remained substantial. 

Together, these patterns suggest a more iterative workflow in which students cycle between interpreting errors, verifying tentative fixes, and implementing solutions. Overall, the distributional changes indicate that student–AI interaction moved from an implementation-driven style (2023), to a verification-driven style (2024), and then to a more balanced debug–verify–implement pattern (2025).

\subsubsection{Prompt Sequence}

\begin{figure}[t]
\centering
\includegraphics[width=\textwidth]{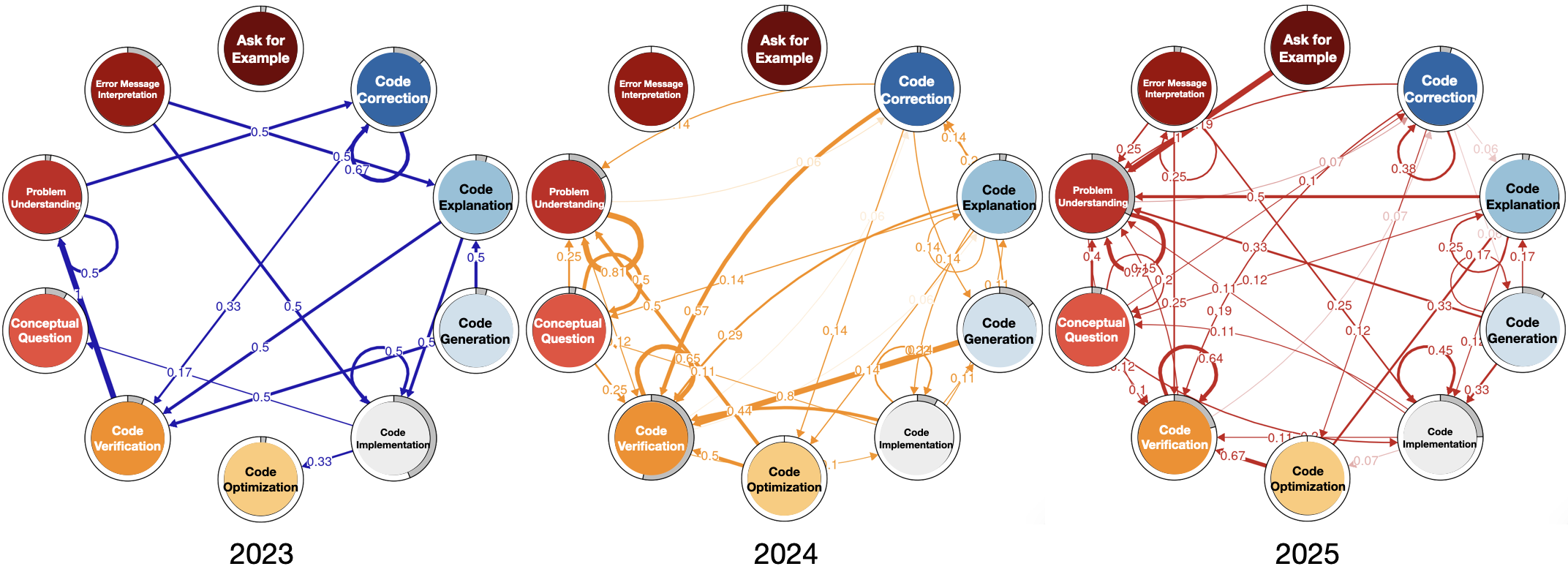}
\caption{Prompt-type transition networks across years.}
\label{prompt_transitions}
\end{figure}

To move beyond “what students asked” and examine “how students progressed,” we analyzed the sequential structure of student prompts using Transition Network Analysis (TNA) \cite{saqr2025transition}. While prompt-type distributions capture the relative prevalence of prompt categories, TNA model the adjacent transitions between prompt types within the same task context, thereby revealing students’ interaction workflows rather than isolated intents. The results of TNA is shown in Fig.~\ref{prompt_transitions}. In the figure, nodes represent the coded prompt types. A directed edge indicates that a prompt of one type was followed by a prompt of another type in the next student message within a task episode; thicker/darker edges denote stronger or more frequent transitions. Self-loops indicate repeated use of the same prompt type across consecutive turns. Thus, network density, hub nodes, and recurrent loops collectively characterize whether students’ use of AI resembles one-shot querying or iterative, multi-step problem solving. 

Across cohorts, the TNAs show a clear structural shift. The 2023 network is comparatively sparse with fewer strong transitions, suggesting shorter and less iterative interaction episodes. In 2024, the network becomes substantially more connected and exhibits a prominent verification-centered structure. By 2025, the network is the most dense and interconnected, indicating broader strategy mixing and more frequent transitions among understanding, implementation, debugging, and verification states.

\textbf{Sparse Transitions and Short Workflows (2023).} The 2023 network contains fewer strong edges overall, with limited cross-type progression. This pattern shows single-purpose or short-lived interactions: students often asked for a specific outcome (e.g., an explanation or a quick check) and then ended the exchange rather than iterating across multiple prompt types. In other words, GenAI use in 2023 more closely resembled a one-shot answer engine than a process-oriented collaborator. A plausible explanation is that 2023 students were still developing basic GenAI literacy, such as how to provide context and create a good prompt. When early responses were incomplete or unhelpful, students may have disengaged quickly rather than pursuing follow-up prompts that would create multi-step workflows. This aligns with the lower interaction frequency observed in 2023 in the weekly interaction analysis.

\textbf{Verification Becomes a Workflow Hub (2024).}
In 2024, the network shows markedly richer connectivity and clearer hubs. Notably, Code Verification emerges as a central organizing state, linking frequently with Code Implementation, Code Correction, and Code Explanation. This structure reflects a recurring iterative pattern that resembles authentic programming practice: students attempt an implementation (or request guidance), verify correctness or code behavior, then revise through correction or request explanations, and verify again. This verification-centered loop suggests that students increasingly used AI not merely to obtain solutions, but to support an iterative debugging and refinement cycle. Importantly, the shift is qualitative: it indicates a change in how students coordinate AI within their problem-solving process, not just an increase in how often they use it.

\textbf{Densest Networks and Greater Strategy Mixing (2025).}
The 2025 network is the most dense and interconnected, with more transitions spanning both “understanding/diagnosis” and “code-action” states. Compared to earlier cohorts, 2025 shows stronger coupling among Problem Understanding and Conceptual Question with implementation-related and debugging-related types (e.g., Error Message Interpretation, Code Correction, Code Explanation, Code Verification). This pattern implies that students more frequently alternated between clarifying intent and concepts, implementing changes, interpreting error signals, revising code, and validating outcomes, i.e., a multi-stage collaborative workflow. Crucially, this is not simply “asking more.” The network structure indicates more sustained back-and-forth progression and broader use of prompt types within the same task episode, consistent with more mature and strategic use of AI as part of the programming workflow.

\begin{table}[t]
\centering
\caption{AI response quality across years.}
\label{ai_quality}
\setlength{\tabcolsep}{17.5pt} 
\renewcommand{\arraystretch}{1.15} 
\begin{tabular}{l r r r}
\toprule
Category & 2023 & 2024 & 2025 \\
\midrule
Correct \& Helpful & 94.67\% & 89.39\% & 91.01\% \\
Correct \& Not Helpful & 5.33\% & 7.82\% & 5.30\% \\
Incorrect \& Helpful & 0.00\% & 0.56\% & 0.00\% \\
Incorrect \& Not Helpful & 0.00\% & 2.23\% & 3.69\% \\
\midrule
Correct (overall) & 100.00\% & 97.21\% & 96.31\% \\
Helpful (overall) & 94.67\% & 89.95\% & 91.01\% \\
\bottomrule
\end{tabular}
\setlength{\tabcolsep}{6pt}
\renewcommand{\arraystretch}{1.0}
\end{table}

\subsubsection{AI Response Quality Across Years}

To assess response quality, we coded sampled GenAI outputs along two dimensions—technical correctness and helpfulness—following an established framework \cite{kazemitabaar2024codeaid,ma2025scaffolding}. 

As shown in Table~\ref{ai_quality}, overall, AI response quality was consistently high across cohorts. The proportion of responses that were both Correct and Helpful ranging from 89.39\% (2024) to 94.67\% (2023), with 91.01\% in 2025. When separating the two dimensions, the overall correctness rate stayed very high, and the overall helpfulness rate was similarly stable. A closer look at the data suggests that the main source of “quality loss” was not widespread incorrectness, but rather misalignment between otherwise-correct responses and students’ immediate needs or course constraints. The share of Correct \& Not Helpful responses was small (5.30–7.82\%) but persistent, consistent with cases where the assistant provided solutions that were technically valid yet too advanced, or insufficiently tailored to the course’s intended approach (e.g., using techniques beyond what was taught, skipping key reasoning steps, or not addressing the exact failure mode). Meanwhile, the Incorrect \& Not Helpful category—though still uncommon—likely reflects a combination of factors, including ambiguous or underspecified prompts (e.g., missing code, input--output examples, or error traces), a lack of course-specific context (e.g., task requirements or constraints), and occasional technical errors in the model’s responses.

\begin{table}[t]
\centering
\caption{Assignment performance and final course outcomes across cohorts.}
\label{performance}
\setlength{\tabcolsep}{12pt}
\renewcommand{\arraystretch}{1.15}
\begin{threeparttable}
\begin{tabular}{l l r r r r}
\toprule
Class &  & 2023\tnote{a} & 2023 & 2024 & 2025 \\
\cmidrule(lr){3-6}
\multicolumn{2}{l}{AI access} & No & Yes & Yes & Yes \\
\midrule
\multirow{2}{*}{Assignment score} 
& Mean & 96.50 & 96.28 & 98.18 & 98.43 \\
& SD   & 8.76  & 7.15  & 3.59  & 4.40  \\
\midrule
\multirow{2}{*}{Final grade} 
& Mean & 4.17 & 4.28 & 4.04 & 4.30 \\
& SD   & 0.57 & 1.02 & 0.57 & 0.86 \\
\bottomrule
\end{tabular}
\begin{tablenotes}\footnotesize
\item[a] Baseline.
\end{tablenotes}
\end{threeparttable}
\end{table}

\subsection{Shift of Performance (RQ3)}
To compare performance across course offerings, we averaged each student’s weekly assignment scores across the semester and examined final course grades at the cohort level. As shown in Table~\ref{performance}, assignment scores were consistently high and tightly clustered across cohorts. A Kruskal--Wallis test found no statistically significant differences across cohorts ($H=3.23$, $p=0.357$). Final-grade means were also broadly similar across years, suggesting that overall performance levels did not shift dramatically despite substantial changes in GenAI adoption and interaction patterns.

Several factors may help explain this stability. First, the weekly assignments were graded under a scheme where students could submit multiple times and performance primarily reflected eventual correctness (i.e., passing all test cases). Second, even in the pre-GenAI baseline offering, students could still seek help through alternative channels (e.g., teaching assistant, peers, or online resources), which likely buffered the marginal effect of GenAI access. Taken together, these results suggest that AI access alone did not substantially change observable performance on course assignments or overall grades.

\section{Conclusion}

This paper reports a three-year classroom study of GenAI use in an introductory Python course. Across cohorts, students entered with steadily higher GenAI familiarity and more routine use. Dialogue logs show a shift from implementation-first, short exchanges to verification-centered, multi-step workflows. Despite these changes in interaction strategies, cohort-level assignment and final-grade outcomes were broadly comparable. These findings provide a longitudinal baseline for future work on designing and evaluating instructional supports that foster effective AI literacy and responsible help-seeking in programming learning. Beyond documenting adoption, we show that as GenAI becomes normalized, students increasingly treat the assistant as a partner for verification and iterative refinement rather than a one-shot code generator. This suggests that the central instructional challenge is not simply whether students use GenAI, but how courses shape productive use and align help-seeking with course expectations. Methodologically, we contribute a classroom-grounded longitudinal evidence base that triangulates surveys, interaction-log coding, and course records to track how GenAI literacy and student--AI interaction patterns evolve over time.

This study has several limitations. First, it focuses on a single institution and course context, so patterns may differ across populations, programming languages, or grading schemes. Second, our cohort-level observational design cannot establish causality (e.g., model improvements or peer norms). Third, dialogue logs capture only in-platform GenAI use and may miss other support channels (e.g., peers or TAs).


\bibliographystyle{splncs04}
\bibliography{base}

\end{document}